\documentclass[12pt]{article}

\usepackage{wrapfig}
\usepackage{array} 
\usepackage{amssymb}
\usepackage{graphics,graphpap}
\usepackage{graphicx}
\usepackage{color}
\usepackage{graphicx}
\usepackage{dcolumn}
\usepackage{epsfig}
\usepackage{epstopdf}
\usepackage{bm}
\usepackage{amsmath}
\usepackage{amsfonts}
\usepackage{bbm}
\usepackage{subfigure}
\usepackage{setspace}

\renewcommand{\thefootnote}{\fnsymbol{footnote}}

\newcommand{\beq}{\begin{eqnarray}}
\newcommand{\eeq}{\end{eqnarray}}
\newcommand{\cw}{{w_T}}
\newcommand{\Cw}[1]{{w_T^{#1}}}
\newcommand{\hoch}{{M_\cU}}
\newcommand{\Hoch}[1]{{M_\cU^{#1}}}

\newcommand{\bmp}{\noindent\begin{minipage}{16cm}}
\newcommand{\emp}{\end{minipage}\vskip 7mm} 

\def\drawbox#1#2{\hrule height#2pt
        \hbox{\vrule width#2pt height#1pt \kern#1pt
              \vrule width#2pt}
              \hrule height#2pt}

\def\Asym#1#2{\vcenter{\vbox{\drawbox{#1}{#2}
              \kern-#2pt 
              \drawbox{#1}{#2}}}}

\def\simge{\mathrel{%
   \rlap{\raise 0.511ex \hbox{$>$}}{\lower 0.511ex \hbox{$\sim$}}}}

\def\simle{\mathrel{
   \rlap{\raise 0.511ex \hbox{$<$}}{\lower 0.511ex \hbox{$\sim$}}}}

\def\s#1{\setbox0=\hbox{$#1$}%
\rlap{\ifdim\wd0>.7em\kern.22\wd0\else\kern.1\wd0\fi /}#1}

\newcommand{\matel}[3]{\langle #1|#2|#3\rangle}
\newcommand{\vev}[1]{\langle #1 \rangle}

\newcommand{\dU}{{d_{\cal U}}}
\newcommand{\cU}{{\cal U}}

\begin{document}

\begin{titlepage}
\begin{flushright}\begin{tabular}{l}
Edinburgh 2008/41 \\ 
IPPP/08/33\\
DCPT/08/66
\end{tabular}
\end{flushright}

\vskip1.5cm
\begin{center}
   {\Large \bf \boldmath   Unparticle \& Higgs as Composites} 
    \vskip1.3cm {\sc Francesco Sannino$^{\,a}$\footnote{sannino@ifk.sdu.dk}  \&
Roman Zwicky$^{\,b\,c}$\footnote{Roman.Zwicky@ed.ac.uk}}
  \vskip0.5cm

$^a$ {\sl HEP Center, University of Southern Denmark, Campusvej 55, DK-5230 Odense M, Denmark.} \\
$^b$ {\sl School of Physics and Astronomy, University of Edinburgh, Scotland} \\
$^c$ {\sl IPPP, Department of Physics, 
University of Durham, Durham DH1 3LE, UK} \, 
\vspace*{1.5mm}
\end{center}

\vskip0.6cm

\begin{abstract}
We propose a generic framework  in which the Higgs and the unparticle
are both composite. The underlying theories are four dimensional, asymptotically free,
nonsupersymmetric gauge theories with fermionic matter.  We sketch a
possible unification of these two sectors at a much higher scale resembling
extended technicolor models. By construction our extensions are natural, meaning
that there is no hierarchy problem. 
The coupling of the unparticle
sector to the composite Higgs  emerges 
as a four-Fermi operator.
The bilinear unparticle operator near the electroweak 
scale has scaling dimension in the range 
$1 < \dU < 3$.
We investigate, in various ways, the breaking of  scale invariance induced by the electroweak scale
resulting in an unparticle condensate.
The latter acts as a natural infrared cut off 
or hadronic scale. 
We give the low-energy effective theory valid near the electroweak scale. The unparticle-Higgs mixing is found to be suppressed within our framework.
\\[0.1cm]

\end{abstract}

\nonumber

\end{titlepage}

\newpage

\setcounter{footnote}{0}
\renewcommand{\thefootnote}{\arabic{footnote}}

It is an exciting possibility that new strong dynamics
could be discovered at the Large Hadron Collider (LHC). The hope is fueled by the fact that some of the best motivated extensions of the Standard Model (SM)  break the electroweak symmetry dynamically \cite{Weinberg:1979bn,Susskind:1978ms}. The new models, passing the precision electroweak tests, are summarized in \cite{Sannino:2008ha}. It is then interesting to explore the possibility to accommodate the unparticle scenario \cite{Georgi:2007ek} into a natural setting featuring  
four dimensional strongly interacting dynamics.

Georgi's original idea is that at high energy there is an ultraviolet (UV) sector coupled to the SM 
through the exchange of messenger fields with a 
large mass scale $\hoch$. Below that scale 
two things happen consecutively. Firstly, the messenger sector decouples,
resulting in contact interactions between the SM and the unparticle sector. Secondly, the latter flows into a non-perturbative infrared (IR) fixed point
at a scale $\Lambda_\cU  \ll \hoch$ hence exhibiting scale invariance;
\begin{equation}
\label{eq:unscen}
{\cal L}   \sim  {\cal O}_{\rm UV} 
  {\cal O}_{\rm SM}    \to  {\cal O}_\cU {\cal O}_{\rm SM} \, .
\end{equation}
The UV unparticle operator is denoted by ${\cal O}_{\rm UV}$ and it posses integer
dimension $d_{\rm UV}$. When the IR fixed point 
is reached the operator ${\cal O}_{\rm IR} \equiv {\cal O}_\cU$ acquires a non-integer scaling dimension $\dU$ through
dimensional transmutation
\begin{equation}
\label{eq:un}
| \matel{0}{{\cal O}_\cU}{P}| \sim (\sqrt{P^2})^{\dU-1}{} \, .
\end{equation}
This defines the matrix element up to a normalization 
factor.
In the regime of exact scale invariance the spectrum of the operator ${\cal O}_\cU$ is 
continuous, does not contain isolated particle
excitations and might be regarded as one of the reasons for the name ``unparticle''. 
The unparticle propagator carries a 
CP-even phase\footnote{The resulting CP violation 
was found to be consistent with the CPT theorem 
\cite{Zwicky:2007vv}.}
\cite{Georgi:2007si,Cheung:2007zza} for space-like 
momentum. 
Effects were found to be most unconventional 
for non-integer scaling dimension $\dU$, e.g.
\cite{Georgi:2007ek,Georgi:2007si} and \cite{Cheung:2008xu}. 

The coupling of the unparticle sector to the SM \eqref{eq:unscen}  breaks the scale invariance
of the unparticle sector at a certain energy. 
Such a possibility was first investigated 
with naive dimensional analysis (NDA)
 in reference
\cite{Fox:2007sy} 
via the Higgs-unparticle coupling of the form
\begin{equation}
\label{eq:H2O}
{\cal L}^{\rm eff}  \sim {\cal O}_\cU |H|^2 \, .
\end{equation}
The dynamical interplay of the unparticle and Higgs sector in connection with the interaction 
\eqref{eq:H2O} has been studied in 
\cite{Delgado:2007dx}.
It was found, for instance, that the
Higgs VEV  induces an unparticle VEV,
which turned out to be  infrared (IR) 
divergent for their assumed range of scaling dimension and forced the authors to introduce 
various IR regulators 
\cite{Delgado:2007dx,Delgado:2008rq}.

{In this work we elevate the unparticle scenario to a natural extension of the SM by proposing a generic framework  in which the Higgs and the unparticle sectors are both \emph{composites} of elementary
fermions.} We use four dimensional, non-supersymmetric asymptotically free gauge theories with fermionic matter.
This framework allows us to address, in principle, 
the dynamics
beyond the use of scale invariance per se.
 
The Higgs sector is replaced by a walking  technicolor model (TC), whereas the unparticle one
corresponds to 
a gauge theory developing a nonperturbative\footnote{
We note that the Banks-Zaks \cite{Banks:1981nn} type IR points, used to illustrate  the unparticle sector in \cite{Georgi:2007ek}, are accessible in perturbation theory. This yields
anomalous dimensions of the gauge singlet  operators which are close to the pertubative ones, resulting in very small  unparticle type effects.} 
IR fixed point (conformal phase)\footnote{Strictly speaking conformal invariance is a larger symmetry than 
scale invariance but we shall use these terms interchangeably
throughout this paper. We refer the reader to reference
\cite{Polchinski:1987dy} for an investigation of the differences.}\footnote{
Only very recently has it been possible to {\it directly} investigate, via lattice simulations, the dynamics of a number of gauge theories\cite{Catterall:2008qk,Catterall:2007yx,DelDebbio:2008wb,DelDebbio:2008zf,Shamir:2008pb,Svetitsky:2008bw,DeGrand:2008dh,Fodor:2008hm,Appelquist:2007hu,Deuzeman:2008sc,Fodor:2008hn}  expected to develop or to be very close an IR fixed point
\cite{Ryttov:2007cx,Ryttov:2007sr,Dietrich:2006cm}. 
The bulk of the lattice results support the theoretical expectations
\cite{Ryttov:2007cx,Dietrich:2006cm}.}. 
By virtue of TC there is no hierarchy problem.
We sketch a possible unification of the two sectors,
embedding the two gauge theories in a higher 
gauge group. The model resembles the ones
of extended technicolor and leads to 
a simple explanation of the interaction between
the Higgs and the unparticle sectors.

The paper is organized as follows. In section 
\ref{sec:scenario} we describe the basic scenario. 
Thereafter
we address the formation of the unparticle VEV 
in section \ref{sec:unVEV} and identify the VEV
as the natural IR cut off in connection with the
dynamical (constituent) fermion mass. 
The comparison with the IR cut off 
suggested by NDA is presented in appendix
\ref{app:NDA}. In section \ref{sec:ETU} we give some
more details about the unified framework. 
The low energy effective Lagrangian, which could also 
be taken as a starting point, is given in section 
\ref{sec:low}. The regularized 
unparticle propagator with IR and UV 
cut off is discussed in section \ref{sec:regProp}. 
The normalization of the unparticle operator 
to our specific model is discussed in appendix \ref{app:BdU}.
In a further section \ref{sec:unmix} we discuss the mixing of
the Higgs with the unparticle based upon the 
previously given effective Lagrangian. 
In appendix \ref{app:heidi} we comment 
on the proposed unparticle limit of the model 
presented in\cite{vanderBij:2007um}.
The paper ends with an outlook in section \ref{sec:outlook}, where possible future directions of research 
in collider physics, lattice and dark matter are discussed.
For example, we put forward the idea of the \emph{Unbaryon} 
as a possible dark matter candidate.

\section{The Higgs \& Unparticle as Composites}
\label{sec:composites}

\subsection{Scenario}
\label{sec:scenario}

\begin{figure}[h]
  \begin{center}
    \includegraphics[width=0.5\textwidth]{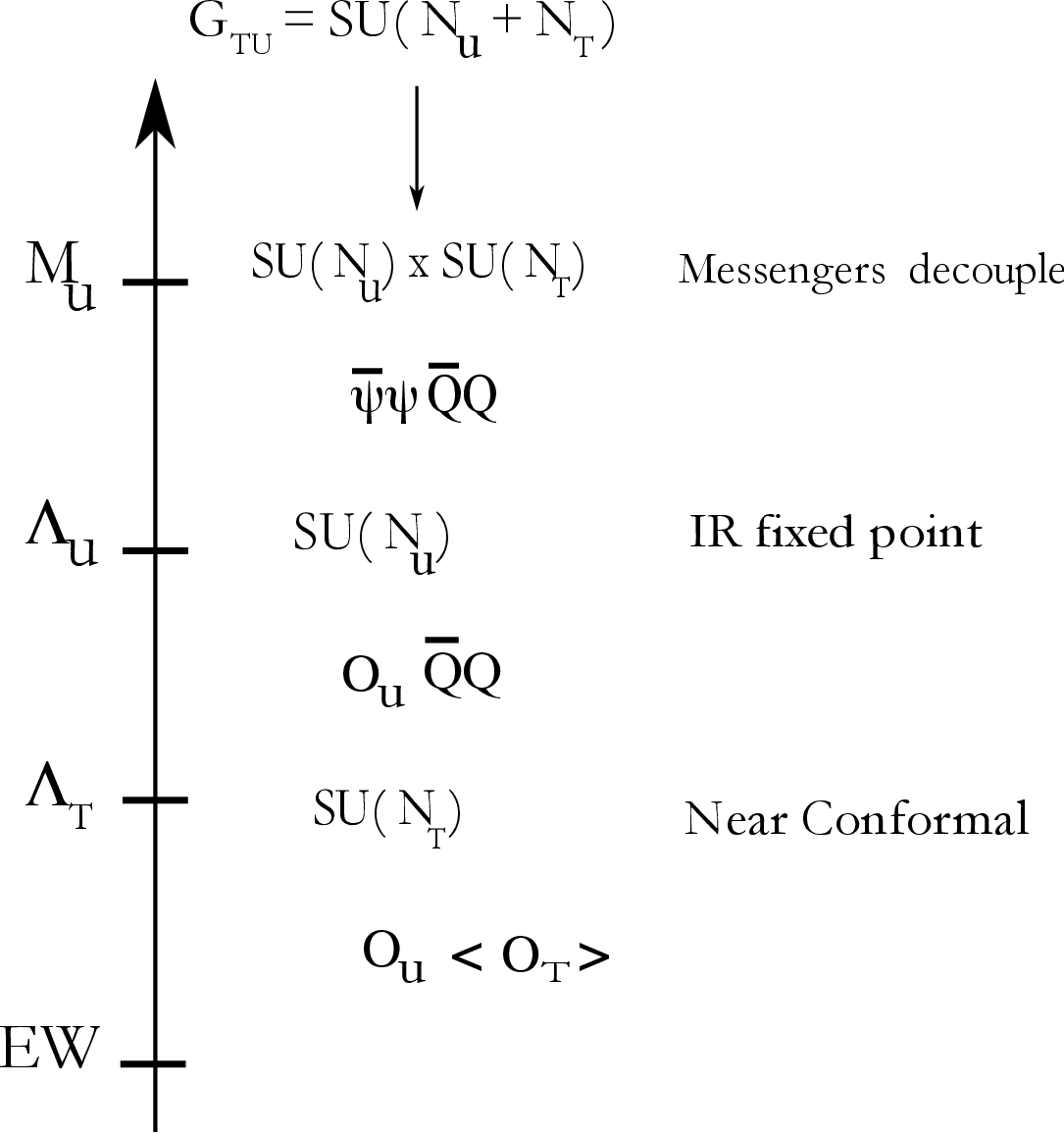}
  \end{center}
  \vspace{-14pt}
  \caption{\small Schematic scenario. The ordering of the
energy scales $\Lambda_\cU$ and $\Lambda_{\cal T}$ is not of any importance.}
\label{fig:unfig}
  \vspace{-5pt}
\end{figure}

Our building block is  an extended 
$G_{T \times U} \equiv SU(N_{T})\times SU(N_{U})$ technicolor (TC)
 gauge theory.
The matter content constitutes of techniquarks  $Q^a_{f}$ charged under
the  representation $R_T$ of the TC group $SU(N_T)$ and 
Dirac techniunparticle fermions  $\Psi^A_{s}$
charged under the representation $R_U$ of the unparticle group 
$SU(N_{U})$, where  
$a/A = 1 \ldots {\rm dim}{[R_{T/U}]}$ and 
$f/s=1 \ldots F/S$ denote gauge and flavor indices
respectively. 
We will first describe the (walking) TC 
and (techni)unparticle sectors separately before
addressing  their common dynamical origin. A graphical illustration of the scenario is depicted 
in Fig.~\ref{fig:unfig} as a guidance for the reader 
throughout this section.

In the TC sector
the number of techniflavors, the matter representation and the number of colors are arranged in such a way that the dynamics is controlled by a near conformal (NC) IR fixed point\footnote{There are a number of ways of achieving (near) conformal dynamics as summarized in \cite{Sannino:2008kg}. 
The state-of-the-art  phase diagram 
\cite{Ryttov:2007sr, Dietrich:2006cm} and
new tools \cite{Ryttov:2007cx}  to construct viable NC nonsupersymmetric gauge theory are reported in \cite{Sannino:2008ha}.}. In this case the gauge coupling reaches almost a fixed point around the scale $\Lambda_{\cal T} \gg M_W$,  with $M_W$ the mass of the electroweak gauge boson.   The TC gauge coupling, at most, gently rises from this energy scale down to the electroweak one.
The coupling is said to {\it walk}\,\footnote{Such models are known as walking TC \cite{Holdom:1981rm}. 
They are preferred 
over QCD-like TC models by the electroweak precision data. In particular, the S-parameter receives a negative contribution for NC models 
\cite{Appelquist:1998xf}. 
A large class of phenomenologically viable models have been identified \cite{Sannino:2004qp,Dietrich:2006cm, Sannino:2008kg,Sannino:2008ha} of which Minimal Walking Technicolor (MWT) and Partially Gauged Technicolor (PGT) constitute two relevant examples.}.
Around the electroweak scale the TC dynamics 
triggers the spontaneous breaking of the electroweak symmetry through the formation of the 
technifermion condensate, which therefore has  the  quantum numbers of the SM Higgs boson. 
The associated Goldstone bosons (technipions) 
then become the longitudinal degrees of freedom of the electroweak bosons in exact formal analogy to the SM.  In the simplest TC models the 
technipion decay constant $F_T$ is related 
to the weak scale as $2 M_W = g  F_T$ ($g$ is the weak coupling constant) and therefore 
$F_{T} \simeq 250$~GeV. The TC scale, analogous 
to $\Lambda_{\rm QCD}$ 
for the strong force, is roughly 
$\Lambda_{TC} \sim 4 \pi F_T$.

Now we turn our attention to the unparticle sector. Here the total number of massless 
techniunparticle flavors $S$ is balanced against the
total number of colors $N_{U}$ in such a way 
that the theory, per se, is asymptotically free and 
admits a nonperturbative IR fixed point.  The energy scale around which the IR fixed point starts to set in is indicated with $\Lambda_{\cal U} \gg M_W$. 

It might be regarded as natural to assume that the unparticle and the TC sectors  have a common dynamical origin, e.g. are part of a larger gauge group at energies  above
$\Lambda_{\cal T}$  and  $\Lambda_{\cal U}$.  
We would like to point out that  
the relative ordering between 
$\Lambda_{\cal T}$ and $\Lambda_{\cal U}$ 
is of no particular relevance for our scenario. The low energy relics of such a unified-type model are four-Fermi operators allowing the two 
sectors to communicate with each other at low
energy.
The unparticle sector will then be driven away from 
the fixed point due to the appearance 
of the electroweak scale in the 
TC sector.

The model, of which further details are presented in section \ref{sec:ETU}
resembles models of extended technicolor (ETC)
\cite{Sannino:2008ha},
where the techniunparticles play the role of the
SM fermions.
We refer to these type of models as 
Extended Techni-Unparticle (ETU) 
models\footnote{The work by Georgi and Kats 
\cite{Georgi:2008pq} on a two dimensional example of unparticles triggered this work.}.  
At very high energies $E \gg M_\cU$ 
the gauge group 
$G_{T \times U}$ is thought to be 
embedded in a simple group 
$G_{\rm TU} \supset  G_{T \times U}$.
At around the scale  $\hoch$ the ETU group 
is broken to $G_{TU} \to 
G_{T \times U } $
and the heavy gauge fields receive masses 
of the order of $\hoch$ and play the role
of the messenger sector. 
Below the scale $\hoch$ the massive 
gauge fields decouple and  four-Fermi operators emerge, which corresponds to the first step
of the scenario, e.g. Eq.~\eqref{eq:unscen} 
and Fig.~\ref{fig:unfig}.
Without committing to the specific ETU dynamics
the interactions can be parametrized as: 
\begin{eqnarray}
\label{eq:ETC}
&&
{\cal L}_{< \hoch}^{\rm eff} = \alpha \frac{\bar{Q}Q \,{\bar{\Psi} \Psi}}{\Hoch{2}} + \beta \frac{{\bar{Q}Q }{\bar{Q}Q }}{\Hoch{2}} + \gamma \frac{{\bar{\Psi}\Psi }{\bar{\Psi} \Psi }}{\Hoch{2}} \,. \label{4fermi}
\end{eqnarray}
The coefficients  $\alpha$, $\beta$ and $\gamma$
(the latter should not be confused with an anomalous dimension) are  of order one, 
which can  be calculated if 
the gauge coupling $g_{\rm TU}$ is perturbative.
The Lagrangian \eqref{eq:ETC} is the relic of the
ETU(ETC) interaction and gives rise to two 
sources of dynamical chiral symmetry 
breaking in addition to the \emph{intrinsic} dynamics 
of the groups $G_{T/U}$.
These are contact interactions of the type emphasized 
in \cite{Grinstein:2008qk}.
Firstly, when one fermion pair acquires a VEV
then the $\alpha$-term turns into a tadpole and 
induces a VEV for the other fermion pair. 
{ This is what happens to the unparticle sector when the TC sector, or the SM Higgs \cite{Delgado:2007dx}, 
breaks the electroweak symmetry. 
Secondly, the $\gamma$ term 
corresponds to a Nambu$-$Jona-Lasino type 
interaction which may lead to the formation of a VEV,
for sufficiently large $\gamma$.
This mechanism 
leads to breaking of scale invariance 
even in the absence of
any other low energy scale.
Let us parenthetically note that this mechanism
is operative in models of top condensation, 
c.f. the TC report \cite{Hill:2002ap} for an overview.
However, based on our analysis in the appendix 
\ref{app:NJL}  
we shall neglect this mechanism in the sequel of 
this paper.
We shall  refer to these two mechanism 
as $\alpha/\gamma$-induced condensates.}

At the scale $\Lambda_\cU \gg  M_W$  the unparticle gauge sector flows into an IR fixed point
and the UV operator 
${\cal O}_{\rm UV} = \bar{\Psi} \Psi$
becomes the \emph{composite unparticle}  operator $
{\cal O}_{\rm IR} \equiv {\cal O}_\cU$  with scaling 
dimension $\dU \equiv 3-\gamma_\cU$\,\footnote{The parametrization $\dU \equiv  3 -\gamma_\cU$ will be standard throughout the entire paper and in the text the scaling dimension $\dU$ and 
the anomalous dimension $\gamma_\cU$ will be used interchangeably.},
\begin{equation}
\label{eq:psis}
(\bar{\Psi} \Psi) _{UV}  \sim \Lambda_{\cal U}^{\gamma_\cU}{\cal O}_{\cU} \; ; \qquad  
M_W\ll \Lambda_{\cU} \ll  \hoch  \ .
\end{equation}
Note, the anomalous dimension $\gamma_\cU$ 
of the operator has to satisfy 
$\gamma_\cU \leq 2$ due 
to unitarity bounds of the representations of the 
conformal group \cite{Mack:1975je}. 
The Lagrangian then simply becomes
\begin{eqnarray}
\label{eq:scem}
{\cal L}^{\rm eff}_{\Lambda_\cU} = 
\alpha' \, \frac{\bar{Q}Q  \,  \Lambda_\cU^{\gamma_\cU} {\cal O}_\cU}{\Hoch{2}} + \beta' \, \frac{\bar{Q}Q \bar{Q}Q}{\Hoch{2}} + 
\gamma' \, \frac{ \Lambda_\cU^{2\gamma_\cU} \, {\cal O}_\cU {\cal O}_\cU}{\Hoch{2}} \, .
\end{eqnarray}
This realizes the second step in the scenario, 
c.f.  Fig.~\ref{fig:unfig} and
Eq.~\eqref{eq:unscen}. {The matching coefficients
$\alpha',\beta',\gamma'$ \eqref{eq:scem} 
are related to $\alpha,\beta,\gamma$ \eqref{eq:ETC} by order one coefficients.} 
The $\alpha$-term in Eq.~\eqref{eq:scem} is
similar to the unparticle-Higgs interaction in Eq.~\eqref{eq:H2O}.

The composite 
operator $\bar{Q}Q$ can be treated 
in analogy  to $\bar \Psi \Psi$ in  
\eqref{eq:psis},
\begin{equation}
\label{eq:tfcond}
(\bar{Q}Q ) _{UV}  \sim \Lambda_{\cal T}^{\gamma_{\cal T}}{\cal O}_{\cal T} \;;  \qquad  
\Lambda_{TC}\ll\Lambda_{\cal T} \ll M_{\cal U} \, ,
\end{equation}
up to  logarithmic corrections which are negligible.
Contrary to the unparticle sector the TC gauge dynamics break scale invariance through the formation of an \emph{intrinsic condensate}
\begin{equation}
\label{eq:TCcond}
\vev{ {\cal O}_{\cal T}}_{\Lambda_{\cal T}}  \simeq
 \Cw{\gamma_{\cal T}}  \Lambda_{TC}^{d_{\cal T}} 
\equiv   \Cw{\gamma_{\cal T}} \Lambda_{TC}^{3 - \gamma_{\cal T}} 
\, , \quad  \cw \equiv \Big(\frac{\Lambda_{\cal T }}{\Lambda_{ TC}} \Big) \ .
\end{equation}
The estimate of the VEV is based on
scaling from QCD and renormalization group evolution.

The relevant terms contained in the low energy effective theory around the electroweak scale 
are\footnote{
Note that in QCD-like TC models 
(the gauge coupling displays a running behavior rather than a walking one) 
one would set  $\gamma_{\cal T} \simeq 0$ in  Eqs.~\eqref{eq:TCcond} and \eqref{eq:OUint}.}:
\begin{equation}
\label{eq:OUint}
{\cal L}_{\Lambda_{\cal T}}^{\rm eff}  =  \alpha' \,  \Cw{\gamma_{\cal T}} \,
\frac{\Lambda_{TC}^{3}  \, \Lambda_\cU^{\gamma_\cU} {\cal O}_\cU}{\Hoch{2}} +  \gamma' \, 
\frac{\Lambda_\cU^{2\gamma_\cU} {\cal O}_\cU 
{\cal O}_\cU}{\Hoch{2}}  + ..   
\end{equation}
This step involves another matching procedure but
we shall not introduce further notation here
and denote the matching coefficients by simple
primes only.
As stated previously the TC condensate 
drives the TC gauge sector away from the
fixed point and the coupling increases towards the IR. The 
sector is then replaced by a low energy effective 
chiral Lagrangian featuring the relevant composite degrees of freedom \cite{Foadi:2007ue,Hill:2002ap}. 
{ The lightest isosinglet composite scalar, 
with $J^{PC} = 0^{++}$ quantum numbers, is the \emph{composite Higgs}. 
For further details c.f. appendix F of reference \cite{Sannino:2008ha} as well as \cite{Hong:2004td}.} 
The linear realization of the chiral Lagrangian of the TC model will be discussed in section \ref{sec:ETU}.


\subsection{Unparticle VEV $\vev{{\cal O}_\cU}$}
\label{sec:unVEV}

We shall now investigate how 
the $\alpha$-term in \eqref{eq:OUint} 
induces a VEV for the unparticle.
In a  theory with canonically normalized fields the mass term,
if present, has to be included for the minimization of the potential.
In the unparticle setup we do not have explicit mass terms but 
there is  a continuous mass spectrum and it is not immediately clear how to proceed.
Yet we can make use 
of the deconstructed version of unparticles, proposed by Stephanov  \cite{Stephanov:2007ry},
in order to imitate the situation of isolated masses \cite{Delgado:2007dx}.
The continuous spectrum is deconstructed 
into an infinite tower of massive particles,
\begin{equation}
\label{eq:dec}
{\cal O}_\cU(x) = \sum_n f_n \varphi_n(x) \, ; \qquad 
\left\{ 
\begin  {array}  {l}   \def  \ma  { \left(  \begin {array} {ccccc} } 
f_n^2 =   \Delta^2  \,  \frac{B_\dU}{2 \pi} (M_n^2)^{\dU-2} \\[0.1cm]
 M_n^2 = n \Delta^2 \\
\end{array}  \right. \,,
\end{equation}
with an adjusted residuum $f_n$ and a finite
spacing $\Delta$ in units of the mass.
The spectral function
$\rho_\dU$ of the operator above is given by
\begin{equation}
\label{eq:adu}
\rho_\dU(P^2)\theta(P_0) = \sum_{n} \delta(P^2 - P_n^2) 
|\matel{0}{{\cal O}_\cU} {P_n}|^2
\quad \rightarrow \quad
\rho_\dU(s) \equiv \frac{B_\dU}{2 \pi} s^{\dU-2} \theta(s)
 \,  .
\end{equation}
The square root of $B_\dU$ defines the strength with
which a state couples to the unparticle operator.
The sum above is a mnemonic for the sum over all the possible states 
in the  Hilbert space. The propagator then follows 
from the spectral function from the 
 K\"all\'en-Lehmann representation \eqref{eq:prop}.
The factor $B_\dU$ corresponds to $A_\dU$ in \cite{Georgi:2007ek}, but we have chosen to denote it by a different letter
since, as we shall see, our model demands a  different form.   
The value $\dU = 2$ in \eqref{eq:adu} is the dividing value between
the IR and UV sensitive domains and  will play a crucial role later on. Other regularizations than the one in Eq.~\eqref{eq:dec} are possible \cite{Stephanov:2007ry}.
Therefore,
no physical interpretation should be attached to it.
The 
Lagrangian \eqref{eq:OUint} with added mass terms becomes 
\begin{equation}
{\cal L } = \bar \alpha \sum_n f_n \varphi_n  + \bar \gamma  \sum_{n,m} f_n f_m \varphi_n \varphi_m  -  \frac{1}{2} \sum_n M_n^2 \varphi_n^2  \, ,
\end{equation}
with 
\begin{equation}
\label{eq:bar}
\bar \alpha \equiv \alpha' w_T^{\gamma_{\cal T}} \frac{\Lambda_{TC}^3 \Lambda_\cU^{\gamma_\cU}}{\Hoch{2}}   \, , \qquad \bar \gamma \equiv \gamma' \frac{\Lambda_\cU^{2 \gamma_\cU}}{\Hoch{2}} \,,
\end{equation}
being the prefactors of 
${\cal O}_\cU$ and ${\cal O}_\cU^2$  in \eqref{eq:OUint}, with mass dimensions
$[\bar \alpha] = \gamma_\cU+ 1 $ and $[\bar \gamma] = 2 \gamma_\cU-2$. The equation of motion (e.o.m.) for the operator 
$\varphi_n$ is
\begin{equation}
 \bar \alpha f_n + 2 \bar \gamma f_n ( \sum _m f_m \varphi_m) -  M_n^2 \varphi_n = 0
\,.
\end{equation}
A simple recursive relation follows from these relations,
\begin{equation}
\label{eq:rec}
\varphi_n = n^{(\dU-4)/2} \varphi_1  \, .
\end{equation}
Inserting this result into the e.o.m. for $\varphi_1$  we obtain, 
\begin{equation}
\label{eq:bdu}
\vev{\varphi_1} = \frac{\bar \alpha \, b_\dU \Delta^{\dU-3}}{ 1 - 2 \bar \gamma (b_\dU \Omega_\Delta)} \,, 
\qquad b_\dU \equiv \sqrt{ \frac{B_\dU}{2 \pi}},
\end{equation} 
where $\Omega_\Delta$ is the sum over the modes,
\begin{eqnarray*}
\Omega_\Delta &\equiv& 
 \Delta^{\dU-3}   \sum_n \Big( \frac{f_n}{b_\dU} \Big) \frac{\varphi_n}{\varphi_1} 
= \Delta^2  \sum_n (n \Delta^2)^{\dU-3}  \,.
\end{eqnarray*}
The  quantities $\vev{\varphi_n}$ are then obtained from the recursion 
relation \eqref{eq:rec} and 
the unparticle VEV is the sum of 
its deconstructed parts \eqref{eq:dec},
\begin{equation}
\label{eq:unVEV}
\vev{{\cal O}_\cU}_\Delta = \sum_n f_n \vev{\varphi_n} =
\frac{\bar \alpha }{1 - 2 \bar \gamma (b_\dU \Omega_\Delta)} (b_\dU \Omega_\Delta) \, . 
\end{equation}
Solving this equation with appropriate UV and IR regularizations 
is the main goal  of the rest of this section.
The unparticle condensate will be connected with the
IR cut off, which implies that Eq.~\eqref{eq:unVEV}
has to be solved in a self consistent way.
Removing the discrete regularization, the 
sum $\Omega_\Delta$ is converted into an integral,
which we shall regularize with an IR and UV regulator for later convenience,
\begin{eqnarray}
\label{eq:Sreg}
\Omega(\Lambda_{\rm IR}, \Lambda_{\rm UV}) = 
\lim_{\Delta \to 0} \Omega_\Delta = 
  \int_{\Lambda^2_{\rm IR}}^{\Lambda^2_{\rm UV}} ds s^{\dU-3}  
 &=&  \Big( \frac{(\Lambda^2_{\rm UV})^{\dU-2} - (\Lambda^2_{\rm IR})^{\dU-2}}{\dU-2} \Big)  \quad .
 \end{eqnarray}
We note that when the quadratic term is removed, 
i.e. $\gamma \to 0$ in \eqref{eq:OUint}, the problem reduces to a single unparticle operator coupled to 
an external source, 
of which the interaction \eqref{eq:H2O} is a special case when the Higgs assumes a VEV. 
In this limit the result in 
Eq.~\eqref{eq:unVEV} indeed reduces to the 
expression found in  reference \cite{Delgado:2007dx}.

The integral $\Omega(\Lambda_{\rm IR}, \Lambda_{\rm UV})$ \eqref{eq:Sreg}  is sensitive
to the UV cut off for $\dU > 2$ and to the 
IR cut off for $\dU < 2$. 
The effective theory for the unparticle operator is
valid up to the scale $\Lambda_\cU$ and is therefore a UV cut off of the theory.
Moreover at energies larger than 
$\Lambda_{\cal T}$ 
there is no $\bar QQ$ condensate, which 
implies $\bar \alpha \to
0$ \eqref{eq:bar} and therefore the modes above
$\Lambda_{\cal T}$ do not contribute to the VEV
in Eq.~\eqref{eq:unVEV}). So effectively 
the UV cut off is the lower of the two scales,
\begin{equation}
\label{eq:UV}
\Lambda_{\rm UV}  \simeq \min(\Lambda_\cU,
\Lambda_{\cal T}) \,.
\end{equation} 
The constituent fermion mass\footnote{
In QCD the condensate induces a dynamical mass, 
the so-called constituent quark mass. 
An estimate can be obtained by extending the definition of the perturbative 
pole mass to include additional terms from the Operator Product Expansion 
\cite{Politzer:1976tv}. Adapting the situation to the case of
a non-trivial fixed point leads to 
$(m_{\rm const})^\dU  \simeq - g_U^2 C_{\dU} \vev{{\cal O}_\cU} $, where $C_\dU$ an order one coefficient which is not calculable due to virulent strong interaction effects. The lowest order QCD result  is recovered by setting $\dU \to  3$ and $C_\dU \to 1$.} $m_{\rm const}$ 
provides a natural IR cut off:
\begin{equation}
\label{eq:IR}
\Lambda_{\rm IR} \simeq 2  m_{\rm const}
\simeq 2 |\vev{{\cal O}_\cU}|^{1/\dU} \ .
\end{equation}For numerical estimates we have chosen the factor two in front of the condensate based on the crude idea of identifying the IR cut off with a possible lightest meson of mass roughly twice the constituent mass. However this choice does not affect the qualitative nature of our results.

\begin{figure}[h]
\centering
 \includegraphics[width=0.48\textwidth]{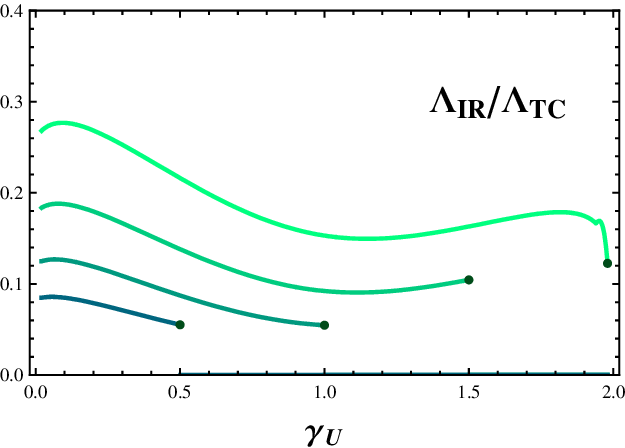}
\includegraphics[width=0.48\textwidth]{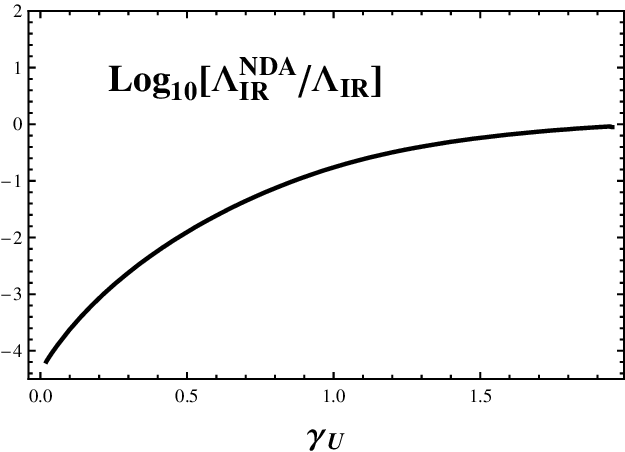}
\caption[]{\small
(left) $\Lambda_{\rm IR}/\Lambda_{TC}$ as a  function 
of $\gamma_\cU$ up to the constraint $\gamma_\cU \leq  \gamma_{\cal T}$.
The actual value of $\gamma_{\cal T}$ can therefore be read-off from the endpoint of the curve. 
(right) Logarithm of the ratio of IR cut offs against 
$\gamma_\cU$ for $\gamma_{\cal T} = 1$. 
The influence of the $\gamma$-term is completely 
negligible for the chosen input values.
The dependence on $\gamma_{\cal T}$ is very mild 
and we have chosen somewhat arbitrarily 
$\gamma_{\cal T} = 3/2$.
Trivial factors, like $b_\dU$, are fixed such that
equality of $\Lambda_{\rm IR}$ and 
 $\Lambda_{\rm IR}^{\rm NDA}$ 
is reached for $\gamma = 0$ 
in the domain $\gamma_\cU = 2$.
In both  figures we have chosen 
$N_U  = 4$ in  the interpolation formula 
\eqref{eq:interpolation}. 
Furthermore the hierarchies of 
scales \eqref{eq:hierarchy} are set to 
$\Lambda_{TC} : (\Lambda_\cU = \Lambda_{\cal T}) : \hoch = 1 : 10^1 : 10^3$ 
and the coefficients $\alpha'$ and $\gamma'$ to a value of unity.}
\label{fig:three}
\end{figure}

Assuming the UV and IR cut offs as in Eqs.~\eqref{eq:UV} and \eqref{eq:IR} respectively the unparticle
condensate can then be extracted from Eq.~\eqref{eq:unVEV} in terms of the scales $\{ \Lambda_{ TC}, \Lambda_\cU, \Lambda_{\cal T}, \hoch \}$.
and the anomalous dimensions $\gamma_\cU$ 
and $\gamma_{\cal T}$. The parameters
obey the following 
hierarchies,
\begin{equation}
\label{eq:hierarchy}
\Lambda_{ TC} \ll \Lambda_\cU , \Lambda_{\cal T} \ll \hoch \,, \qquad  \gamma_\cU  < \gamma_{\cal T} \,,
\end{equation}
c.f. Fig.~\ref{fig:unfig} for the scales and 
section \ref{sec:ETU} for an explanation concerning
the relation of the anomalous dimensions.
{ We investigate 
Eq.~\eqref{eq:unVEV} analytically 
in the follwing 
three  regimes }
\vspace{0.2cm}
\begin{alignat}{3}
\label{eq:eq1}
& {\rm Approximate \; solution \;of\; Eq.}\mbox{\eqref{eq:unVEV}} \;{\rm for}\; \vev{\hat{\cal O}_\cU}& & {\rm valid\; near} & & {\rm sensitive} \nonumber \\
&  \vev{\hat {\cal O}_\cU} (|\vev{\hat {\cal O}_\cU}|^{ \frac{2\gamma_\cU-2}{3-\gamma_\cU}} -      2 C_1 \gamma') -   C_0  \alpha'  = 0  & & 
\left\{
\begin  {array}  {l}   \def  \ma  { \left(  \begin {array} {ccccc} } 
 \dU \gtrsim  1  \\
\gamma_\cU \lesssim 2
\end{array}  \right.
 \quad  & &  {\rm IR}
\\
& \; \; C_1 =   \frac{b_\dU  }{\gamma_\cU-1}  \Big(  \frac{\Lambda_\cU^2}{\Hoch{2}} \Big) \Big(  \frac{(2\Lambda_{TC})^2}{\Lambda_{\cU}^2} \Big)^{1-\gamma_\cU} 
\quad C_0 = C_1 \Big(  \frac{\Lambda_{TC}}{\Lambda_\cU} \Big)^{\gamma_\cU}
\Cw{\gamma_{\cal T}} & & & & 
\nonumber  \\[0.1cm]
 \label{eq:eq2}
& \vev{\hat  {\cal O}_\cU} \simeq - \alpha' b_2 \Big(  \frac{\Lambda_{\cU} \Lambda_{TC}}{\Hoch{2}} \Big) \Cw{\gamma_{\cal T}} \log \Big[  \frac{ \Lambda_{TC}^2}{\Lambda_\cU^2} 4 |\vev{\hat  {\cal O}_\cU}| \Big]
 \, \qquad  & & \left\{
\begin  {array}  {l}   \def  \ma  { \left(  \begin {array} {ccccc} } 
 \dU = 2  \\
\gamma_\cU = 1
\end{array}  \right.
 \quad  & &  {\rm IR \sim UV} \\[0.1cm]
 \label{eq:eq3}
& \vev{\hat  {\cal O}_\cU} \simeq 
+  \alpha'  
\frac{b_\dU}{1-\gamma_\cU}  \,
\Big(  \frac{\Lambda_\cU^2}{\Hoch{2}} \Big) \Big(  \frac{\Lambda_{TC}^2}{\Lambda_\cU^2} \Big)^{\gamma_\cU}
\Cw{\gamma_{\cal T}}  \, \qquad  & &  \left\{
\begin  {array}  {l}   \def  \ma  { \left(  \begin {array} {ccccc} } 
 \dU   \lesssim  3  \\
\gamma_\cU \gtrsim    0 
\end{array}  \right.
 \quad  & &  {\rm UV} \, \qquad 
\end{alignat}
We have chosen to normalize the unparticle VEV,
\begin{equation*}
\vev{\hat { {\cal O} }_\cU} \equiv  
\frac{\vev{  {\cal O}_\cU}}{\Lambda_{TC}^\dU}  \,,
\end{equation*}
to  the chiral symmetry breaking scale of the 
TC sector. 
Eqs.~\eqref{eq:eq1} and \eqref{eq:eq3} are sensitive to the IR and UV domain
of \eqref{eq:Sreg}. 
The solutions are valid in a small neighborhood of $\dU \gtrsim 1$ 
and $\dU \lesssim 3$ respectively.
Note, the UV sensitive domain $\dU \lesssim 3$ 
corresponds to a perturtbative Banks-Zaks type fixed point \cite{Banks:1981nn}.
Eq.~\eqref{eq:eq2} represents the domain which is equally sensitive 
to the IR and UV. We have set $\dU = 2$ strictly for presentational convenience only.
The $\gamma$-term is solely important for 
\eqref{eq:eq1} or more precisely is of the same order as the $\alpha$-term for typical values of the model parameters.
For  $\alpha',\gamma' \geq 0$ all solutions are positive.
In Fig.~\ref{fig:three}(left) we have plotted the IR cut off \eqref{eq:IR} as a function 
of $\gamma_\cU$ for different $\gamma_{\cal T}$ up to the bound $\gamma_\cU \leq  \gamma_{\cal T}$ 
\eqref{eq:hierarchy}. The input values, which are thought to be typical, are indicated in the caption.

The breaking of scale invariance, due to the 
coupling to the Higgs sector,
was investigated in an earlier reference by the use 
of  na\"ive dimensional analysis (NDA) \cite{Fox:2007sy} .
In the appendix \ref{app:NDA} we compare their results
with ours.  At the parametric level we find,
\begin{equation}
\label{eq:IRcomp}
\Lambda_{\rm IR}  \gtrsim     \Lambda^{\rm NDA}_{\rm IR} \, ,
\end{equation}
c.f. Fig.~\ref{fig:three}~(right).
The difference being caused by the fact that in the NDA 
analysis it is implicitly assumed that the unparticle sector
scales with the IR cut off whereas in our model the 
unparticle condensate can also be sensitive to the
UV domain. 
Parametric equality is reached in the region of IR sensitivity,
e.g. \eqref{eq:eq1}.
Needless to say that with NDA factors of $4\pi$ can go unnoticed. In connection with the latter
a similar criticism could apply to our prescription in \eqref{eq:IR}. Nevertheless it appears to us that it is
physically motivated and to some extent is backed up 
from our empirical knowledge of QCD.

So far we have not specified the normalization factor $B_\dU$ introduced in 
Eqs.~\eqref{eq:dec} and 
\eqref{eq:adu}. In appendix \ref{app:BdU} we motivate the following  formula
\begin{equation}
\label{eq:interpolation}
(B_\dU)_{\rm interpol} = 2 \pi (\dU -1) + \Big( \frac{N_U}{16 \pi}-\pi
\Big) (\dU - 1)^2 \, ,
\end{equation}
as interpolation formula between the value of $B_3$, which is determined by the
free fermion loop in our model,  and the behaviour around $B_1$, which is model independent
\cite{Mack:1975je}. As previously stated it differs from the normalization factor $A_\dU$ in 
reference \cite{Georgi:2007ek}.

In the next section we will discuss an ETU model in some more detail.

\section{A Schematic ETU Model}
\label{sec:ETU}

We imagine that at an energy much higher than the electroweak scale the theory is described by a 
 gauge theory
 \begin{equation}
 {\cal L}^{\rm UV} = - \frac{1}{2} {\rm Tr}\left[{\cal F}_{\mu\nu}{\cal F}^{\mu\nu} \right] +  \sum_{{\cal F} = 1}^{F} 
 \bar \xi_{\cal F} (i \s{\partial} + g_{TU} \s{\cal A} )\xi_{\cal F} + .. \,
 \label{UV}
 \end{equation}
where ${\cal A}$ is the gauge field of the  
$SU(N_{T}  +  N_U)$ group and gauge indices
are suppressed.  
$({\xi_{\cal F}^{{\cal A}}})^T = (Q^1 ...  Q^{N_{T}}, \Psi^1 ... \Psi^{N_U} )_{\cal F}$ is the fermion field unifying the technifermion and TC matter content.  The dots in (\ref{UV}) stand for the $SU(3)\times SU(2)_L \times U(1)_Y$ gauge fields and 
their interactions to the SM fermions and technifermions. There is no elementary Higgs field in this formulation.  Unification of the TC and techniunparticle dynamics, as outlined in section \ref{sec:composites}, constrains the flavor symmetry of the two sectors to be identical at high energies. The matter content and the number of technifermions (TC + techniunparticles) is chosen, within the phase diagram in \cite{Sannino:2008ha}, such that the theory is asymptotically free at high energies. The non-abelian global flavor 
symmetry is  $SU_L(F)\times SU_R(F)$. 

At an intermediate scale $M_{\cal U}$, much higher than the scale where the unparticle and TC subgroup become strongly coupled, the dynamics is such that $SU(N_{T} \!+\! N_U)$ breaks to $SU(N_{T})\times SU(N_U)$. Only two flavors { (i.e. one electroweak doublet)} are gauged under the electroweak group. The global symmetry group breaks explicitly to $G_F = SU_L(2)\times SU_R(2) \times SU_L(F-2)\times SU_R(F-2)$. At this energy scale the weak interactions are, however, negligible and we can safely ignore it. 

At the scale $M_{\cal U}$ there are the $Q^i_c$ fermions  - with $i=1,\ldots, F$ and $c=1,\dots, N_T $ - as well as the $\Psi^i_u$ ones - with $i=1,\ldots, F$ and $u=1,\dots, N_U $. Assigning the indices
$i=1,2$ to the fermions gauged under the electroweak group we observe that not only the TC fermions are gauged under the electroweak but also the technunparticles. { To ensure that the 
unparticle sector is experimentally not too visible we have to 
assume a mechanism 
that provides a large mass to the charged techniunparticle fermions.} In reality this is quite a difficult task, since we do not want to break the SM weak symmetry explicitly\footnote{One could for instance 
unify the flavor symmetry of the unparticles with the technicolor gauge group into an
ETC group.
This would also produce a Lagrangian of the type \eqref{4fermi}.
The TC fermions would 
be charged under the electroweak group separately.}.
Our treatment below, however, is sufficiently general to be straightforwardly adapted to various model constructions.

As already stated in the first section, the number of
flavors and colors for the 
TC and unparticle gauge groups 
$SU(N_T)$  and $SU(N_U)$
have to be arranged such that the former is 
NC and the latter is conformal. This enforces the 
conditions:
\begin{equation}
F \leq F^{\ast}_{N_T},  \qquad  F^{\ast}_{N_U}  \leq F  -2 \ .
\label{ftc}
\end{equation}
$F^{\ast}_{N}$ denotes the critical number of flavors, for a given number of colors $N$, 
above which the theory develops an IR fixed point. Recall that two unparticle
flavors are decoupled and hence $F \to F-2$ in the
second inequality in \eqref{ftc}.

 According to the conjectured all order beta function \cite{Ryttov:2007cx} $F^*_N$ is 
 \begin{equation}
 F^{\ast}_N = \frac{11N}{\gamma^{\ast} +2} \ ,
\label{basic}
\end{equation} 
for an SU(N) gauge theory with matter in the fundamental representation. This restricts 
$F^*_N$ inasmuch as the critical anomalous 
dimension has to satisfy the unitarity bound 
$\gamma^{\ast}\leq 2$ \cite{Mack:1975je}.
The ladder approximation, for instance, yields $\gamma^{\ast} \approx 1$  \cite{Appelquist:1988yc}. Combining Eqs.~(\ref{ftc}) and (\ref{basic}) we 
arrive at the following allowed  window for the number of flavors:
\begin{equation}
\frac{11N_U}{\gamma^{\ast} + 2} + 2 \leq F \leq \frac{11N_T}{\gamma^{\ast} + 2} \ .
\label{windowF}
\end{equation}
The anomalous dimension of the mass operator for the unparticle and TC fermions at the fixed point are
\begin{equation}
\gamma_{\cal U } = \frac{11N_U - 2 F + 4}{F-2} \ , \quad  \quad \gamma_{\cal T } = \frac{11N_T - 2 F}{F} \ .
\end{equation}
{}They follow from the conjectured all order beta function
\cite{Ryttov:2007cx}.
For walking TC $\gamma_{\cal T}$ is, in fact, very near $\gamma^{\ast}$ and $F$ is very close to the upper bound of equation (\ref{windowF}). Conformality of the unparticle sector requires  $\gamma_\cU$ to be smaller than $\gamma^*$. 
Summarizing:
\begin{equation}
\label{eq:anom}
 \gamma_{\cal U}   \,< \,  \gamma^{\ast} 
\,\lesssim\, \gamma_{T}  \, .
 \end{equation}
  
\subsection{Low energy description}
\label{sec:low}

{ Below the scale $M_{\cal U}$ all 
four-Fermi interactions have to respect
the  flavor symmetry $G_F$.} The most general four-Fermi operators  have been classified in \cite{Appelquist:1984rr} and the coefficient of the various operators depend on the specific model used to break the unified gauge theory. Upon Fierz rearrangement, the operators of greatest phenomenological relevance are, 
\begin{equation}
{\cal L}^{\rm eff} = \Big( \frac{G}{2} \bar{\Psi}_L \Sigma \Psi_R  +{\rm h.c.}\Big)+    \frac{G^{\prime}}{2M_{\cal U}^2} (\bar{\Psi}_L \Psi_R )(\bar{\Psi}_R \Psi_L) + \dots  \ ,
\end{equation}
the scalar-scalar interactions of Eq.~\eqref{4fermi}.
Here $\Sigma$ is the quark bilinear,
\begin{equation}
\Sigma_i^j  \sim ( {{Q}_L}_i {\bar{Q}_R^j})_{ UV}  \ , \qquad i=1,\ldots , F\ .
\end{equation} 
The flavor indices are contracted and the sum starts from the index value $3$;
the first two indices correspond to the $\Psi$'s charged under the electroweak force, which are decoupled at low energy. 
The fermion bilinear becomes the unparticle operator
\eqref{eq:tfcond},
\begin{equation}
\big( {\cal O_{\cal U}}\big)_i^{\;j}  =  \frac{{{\Psi}_L}_i \bar{\Psi}_R^j}{\Lambda_{\cal U}^{\gamma_{\cal U}}} \ . 
\end{equation}
The matrix $\Sigma$ at energies near the electroweak symmetry breaking scale is identified with the interpolating field for the mesonic composite operators.

{ To investigate  the coupling to the composite Higgs  we write down the low energy effective theory using linear realizations.  We parameterize 
the complex $F \times F$ matrix $\Sigma$ by
 \begin{equation}
 \Sigma = \frac{\sigma + i\,\Theta}{\sqrt{F}} + \sqrt{2}(i \Pi^a + \widetilde{\Pi}^a) {T^a} \ , 
 \end{equation}
where ($\sigma$,$\tilde \Pi$) and
($\Theta$,$\Pi$) have 
$0^{++}$ and $0^{-+}$ quantum 
numbers respectively.
The Lagrangian is given by
\begin{eqnarray}
\label{eq:sigmaeff}
{\cal L}^{\rm eff}  & = &   \frac{1}{2}{\rm Tr} \left[ (D\Sigma)^{\dagger} D\Sigma\right]  - k_1( {\rm \widehat{Tr}}\left[ \Sigma^{\dagger}\cal O_{\cal U}\right]+ {\rm h.c.}) - k_2 {\rm \widehat{Tr}}\left[ \cal O_{\cal U}  {\cal O_{\cal U}}^{\dagger} \right] \nonumber \\ &&  -  m^2_{ETC}\sum_{a=4}^{F^2-1}\frac{\Pi^a\Pi^a}{2}  - V(\Sigma, \Sigma^{\dagger}) \ ,
\end{eqnarray}
where
\begin{eqnarray}
D\Sigma = \partial \Sigma - ig W\Sigma + ig^{\prime}\Sigma BT^3_R \ ,  \quad {\rm and} \quad  W=W^a T^a_L  \ , 
\end{eqnarray}
and $ {\rm Tr}[T^a_{L/R} T^b_{L/R}] = \delta ^{ab}/2$.
The coefficients $k_1$ and $k_2$ are directly proportional to the  $\alpha^{\prime}$ and $\gamma^{\prime}$ coefficients in (\ref{eq:scem}).
The $hat$ on some of the traces indicates that the summation is only on the flavor indices from $3$ to $F$. 
Three of the Goldstone bosons play the role of the longitudinal gauge bosons and the remaining 
ones receive a mass $m^2_{ETC}$
from an ETC mechanism.
We refer the reader to reference \cite{Hill:2002ap} for discussion of different ETC models with 
mechanisms for sufficiently large mass generation.
The first term in the Lagrangian is responsible for the mass of the weak gauge bosons and 
the kinetic term for the remaining Goldstone bosons. 
The VEV's for the flavor-diagonal part of the unparticle operator, reduces to the computation performed in the previous section. 
The potential term preserves the global flavor symmetry $G_F$.
Up to dimension four, including the determinant responsible for the $\eta'$ mass in QCD, the terms  respecting the global symmetries of the TC theory are:
\begin{equation}
V(\Sigma, \Sigma^{\dagger}) = -\frac{m^2}{2} {\rm Tr} \left[ \Sigma^{\dagger}\Sigma\right] + \frac{ \lambda_1}{F} {\rm Tr} \left[ \Sigma^{\dagger} \Sigma\right]^2 + {\lambda_2}
 {\rm Tr}[( \Sigma^{\dagger}\Sigma) ^2 ] - \lambda_3( \, {\rm det}\Sigma +
  {\rm det}\Sigma^\dagger ) \ .\end{equation} 
The coefficient $m^2$ is positive to ensure chiral symmetry breaking in the TC sector. The Higgs VEV enters as follows,
\begin{equation}
\sigma = v + h \ , \qquad  {\rm with} \qquad  F_T = \sqrt{\frac{2}{F}} \, v \simeq 250 \, {\rm GeV} \ .
\end{equation}   
$F$ here is the number of flavors and $h$ the composite field with the same quantum numbers as the SM Higgs. The particles $\sigma,\Theta, \tilde \Pi$ 
all have masses of the order of $v$. The Higgs mass, the Higgs VEV and the $\Theta$ mass, for instance, are} 
\begin{eqnarray}
v^2 = \frac{m^2}{(\lambda_1 + \lambda_2 - \lambda_3)} \,,\quad 
m_h^2 =  2 m^2 \,, \qquad m_\Theta^2 = 4 v^2 \lambda_3^2
\, ,
\end{eqnarray} 
up to corrections of the order of $O(\Lambda_{TC}^2/\Hoch{2})$
due to contributions from $\alpha$-terms.

The lightest pesudoscalars of the unparticle sector
are the pseudo  Goldstone bosons
emerging from the explicit breaking of the global flavor 
symmetry in the unparticle sector. Their mass
can be read off from the linear term 
in $O_\cU$ of the effective Largrangian
\eqref{eq:sigmaeff} 
\begin{equation}
m_{ \Pi_\cU }^2 \simeq \Lambda_{TC}^2 
\left(  \frac{\Lambda_{TC}}{\Hoch{2}} \right)^2 
\left(  \frac{\Lambda_\cU}{\Lambda_{TC}} \right)^{\gamma_\cU} 
\left(  \frac{\Lambda_{\cal T}}{\Lambda_{TC}} \right)^{\gamma_{\cal T}}  \, .
\end{equation}

\subsection{Regularized unparticle propagator}
\label{sec:regProp} 
 
In our model the unparticle propagator 
to be used for phenomenology, defined from the  
the K\"all\'en-Lehmann representation 
\eqref{eq:prop}, is 
\begin{equation}
\label{eq:prop2}
\Delta_\cU(q^2,\Lambda_{\rm UV}^2,\Lambda_{\rm IR}^2) = 
- \frac{B_\dU}{2 \pi} \int_{\Lambda_{\rm IR}^2}^{\Lambda_{\rm UV}^2} 
\frac{ds\, s^{\dU-2}}{s - q^2- i0}  +   s.t.    \; .
\end{equation}
For $\dU > 2$ the integral is sensitive to 
the UV completion, of which the subtraction terms 
$(s.t.)$ are a mnemonic. More precisely,  the part which 
is sensitive to $\Lambda_{\rm UV}$ is ambiguous 
due to the presence of, in principle computable, counterterms, which are expected to be of order one\footnote{The counterterms are expected to be of order one in a theory
which is not fine tuned. This is also known under
the term: 'naturalness'. In our model the UV completion 
is known and the counterterms could in principle be determined, but in practice this is outside the scope of our possibilities.}.
This will limit, in practice, the predictivity 
of the theory .
Modeling 
the UV and IR transition regions by  hard cut offs 
is of course a crude model.
Yet this should not be relevant as long
as $q^2$ is sufficiently far away from these cut offs.
Whereas, for $q^2$ close to the cut off, the integral has an 
endpoint singularity which is, to a great extent, a model
artefact. The situation could be ameliorated for instance
by smearing the momentum with a smooth probability 
density.
Due to the breaking of the scale invariance there
will be single and multiparticle states appearing
in the spectrum, which will affect the $q^2 \sim 
\Lambda_{\rm IR}^2$ behavior.
Having made these statements, we now turn
to the evaluation of the integral in \eqref{eq:prop2}. 
It can be expressed as the  difference  of an IR and UV part, 
\begin{equation}
\label{eq:regprop}
\Delta_\cU( q^2,\Lambda_{\rm UV}^2,\Lambda_{\rm IR}^2) = 
f_\dU(\Lambda_{\rm IR}^2, q^2+i0) - f_\dU(\Lambda_{\rm UV}^2, q^2+i0)\,,
\end{equation}
given by,
\begin{equation}
f_\dU(\Lambda^2,q^2) \equiv  
 \Big[ \frac{B_\dU}{2\pi} \frac{(\Lambda^2)^{\dU-2}}{\dU-2} \Big] \underbrace{\!\! \phantom{x}_2F_1(1,2-\dU,3-\dU,\frac{q^2}{\Lambda^2})  }_{\equiv   \bar f_\dU(q^2/\Lambda^2 ) }\, .
\end{equation}
For later convenience we give 
the behaviour of the function $\bar f_\dU(x)$ for small
and large argument appropriate for the respective domains:
\begin{equation}
\label{eq:as}
\bar f_\dU(x+i0)  =
\left\{  \begin  {array}  {ll}   \def  \ma  { \left(  \begin {array} {ccc} } 
a_0 + a_1 x + O(x^2) & \quad  x \ll 1 \quad {\rm UV} \\[0.1cm]
a_{\dU-2} (-x-i0)^{\dU-2}  + a_{-1} \frac{1}{x} + O(\frac{1}{x^2}) & \quad  x \gg 1 \quad {\rm IR}
\end{array}  \right. \,,
\end{equation}
where the leading coefficients are given by
\begin{alignat}{2}
\label{eq:anum}
a_0 &= 1 \,, \qquad \qquad  \quad   & a_{-1} &= - \frac{\Gamma(1-\dU) \Gamma(3-\dU)}{\Gamma(2-\dU)^2} 
\nonumber \\[0.1cm]
a_1& =  \frac{\dU-2}{\dU-3}  \,, \quad  &a_{\dU-2} &= 
\Gamma(3-\dU)\Gamma(\dU-1) \, .
\end{alignat}

\subsubsection{IR region: $ 1 < \dU < 2$}
In the domain $ 1 < \dU < 2$ the regularized
propagator is  close to the 
propagator 
without IR and UV regularization presented in 
\cite{Georgi:2007si}. 
From the expansion \eqref{eq:as} one immediately obtains,
\begin{equation}
\label{eq:georgi}
\lim_{\begin{array}{c}
\Lambda_{\rm IR} \to 0 \,
\Lambda_{\rm UV} \to \infty 
\end{array}} \Delta_\cU( q^2,\Lambda_{\rm UV}^2,\Lambda_{\rm IR}^2) = \frac{B_\dU}{2 \sin(\dU \pi)} (-q^2-i0)^{\dU-2}  \, \quad 1 < \dU < 2 \, ,
\end{equation}
using $\Gamma(z)\Gamma(1-z)\sin(\pi z)= \pi$.
Note that for finite cut offs the UV part of the propagator
is suppressed by 
$(\Lambda_{\rm IR}/\Lambda_{\rm UV})^{2(2-\dU)}$ and is therefore of minor importance for $\dU$ close to 
$1$.

\subsubsection{UV region: $2 \leq \dU < 3$}
\label{sec:UV}

As previously stated, for $\dU > 2$ the UV part becomes increasingly dominant and  manifests itself in the
appearance of counterterms.
In fact in the strict limit $\dU \to 3$, for example,
the UV contribution is formally the same as the fermion loop contribution to the Higgs mass 
in the SM, e.g. Fig.~\ref{fig:bubble}, which is quadratically divergent.  {
The effective theory is valid for
$q^2 \ll \Lambda_{\rm UV}^2$ and therefore the coefficient $a_0$ \eqref{eq:as} 
is relevant for the UV part of the propagator
\eqref{eq:regprop}. In practice this means that
only a single counterterm, the one
associated with $a_0$ is relevant. As stated earlier, by naturalness the counterterm are expected to be 
of order one.}

Note that 
the limit $\dU \to 2$ leads to a particularly simple expression
\begin{equation}
\lim_{
\dU \to 2 } \Delta_\cU( q^2,\Lambda_{\rm UV}^2,\Lambda_{\rm IR}^2) = \frac{B_\dU}{2 \pi} \log\left(\frac{-q^2-i0}{\Lambda_{\rm UV}^2}\right)  + s.t. \, ,
\end{equation}
where $s.t.$ stands for subtraction terms 
(counterterms).

\subsection{Unparticle-Higgs mixing}
\label{sec:unmix}

We shall now turn to the question of the mixing of the 
unparticle and the Higgs. 
{ Our findings resemble results from extra dimensional
models. E.g. the model called HEIDI  \cite{van der Bij:2006pg}, where the continuous spectrum is mimiked by an infinite tower of narrowly spaced Kaluza-Klein modes. The difference is that our model is inherently four dimensional and that the parameters, such as the IR cut off and the strength of the unparticle-Higgs coupling, are related to each other.
Our model is also different from the one in reference \cite{Delgado:2007dx} since, although both are in four dimensions, the Higgs and unparticle coupling emerges dynamically within a UV complete theory. }

The interaction term from Eq.~\eqref{eq:sigmaeff}
\begin{equation}
\label{eq:Lmix}
 {\cal L}^{\rm mix} = - g_{h {\cal O}_\cU} \, h {\cal O}_
 \cU    \,, \qquad 
g_{h {\cal O}_\cU} = \frac{k_1(F-2) }{\sqrt{F}} \,,
\end{equation}
introduces a mixing between the Higgs and the unparticle. The constant $k_1$ has mass dimension 
$\gamma_\cU$. Its size, on which  we will comment below, is crucial for the qualitative nature of the
physics. 
The Higgs propagator is obtained from inverting
the combined Higgs-unparticle system
\begin{equation}
\label{eq:higgsprop}
\Delta_{hh}(q^2) = \frac{1}{q^2-m_h^2 - g_{h{\cal O}_\cU}^2 
\Delta_\cU(q^2,\Lambda_{\rm UV}^2,\Lambda_{\rm IR}^2)} \, .
\end{equation}
This, of course, results in unparticle corrections controlled by $g_{h{\cal O}_\cU}$. The propagator
can be rewritten in terms of a dispersion representation 
\begin{equation}
\label{eq:higgsdisperse}
\Delta_{hh}(q^2) = - \int \frac{ds\,\rho_{hh}(s)}{s - q^2- i0}  \,,
\end{equation}
where the density, 
\begin{equation}
\label{eq:norm}
\int ds \rho_{hh}(s) = 1 \, ,
\end{equation}
is automatically normalized to unity.
The non zero value of the coupling $g_{h {\cal O}_\cU}$  results solely 
in a change of basis (or poles and cuts) 
of the intermediate particles but does not change
the overall density of states.
A direct way to derive  
\eqref{eq:norm} is to equate the representations 
\eqref{eq:higgsdisperse} and \eqref{eq:higgsprop}, multiply them by  $q^2$ and take the limit $q^2 \to \infty$ resulting in \eqref{eq:norm}.
Please note, 
this only works in the case where $\dU < 3$, 
for which the interaction \eqref{eq:Lmix} 
is power counting renormalizable.
If this condition is not fulfilled one could resort
to a subtracted dispersion relation.

The dispersion representation can be split into 
resonance\footnote{The pole description is only adequate in the narrow width approimation. The Higgs width 
is of course rather sizable in a theory of strong interactions. The presentation 
below is meant to be for illustrative purposes 
only.} 
and  continuum contributions,
\begin{equation}
\rho_{hh}(s) =  \sum_i r_i \delta(s-\bar m_i^2) + \sigma(s) \, .
\end{equation}
The resonance contribution, if present, 
can then be obtained from the pole equation
\begin{equation}
\label{eq:pole}
\Delta_{hh}^{-1}(\bar m_i^2) = 0 \,, \qquad 
r_i  = \left| \frac{d \, \Delta_{hh}^{-1}(s)}{ds}\right|^{-1}_{s=\bar m_i^2} \leq 1 \, .
\end{equation}
The residues $r_i$ are smaller (or equal) to one as
a consequence of the normalization condition 
Eq.~\eqref{eq:norm}.
The continuum is simply given by the cut,
\begin{equation}
\label{eq:cont}
\sigma(s) = \Theta(s- \Lambda_{\rm IR}^2)  {\rm Im}[\Delta_{hh}^{-1}(s)] \, ,
\end{equation}
which corresponds to the imaginary part; most familiar from the optical theorem.

To a large extent the spectral function is
characterized by the zeros of the pole
equation and the onset of the continuum 
relative to the poles. This will depend on 
the strength of the mixing and the anomalous dimension. Somewhat exotic effects can be obtained
when the mixing term is made very large 
\cite{Delgado:2008px,vanderBij:2007um}\footnote{It should also be mentioned that for very large mixing
the theory typically becomes unstable.
The pole equation has tachyonic solutions
and vertices grow in an uncontrolled manner, 
indicating the appearance of a new vacuum.
It is possible though that interesting effects
could arise in a somewhat intermediate regime.}. 
In our model the mixing is determined by
$k_1$ \eqref{eq:Lmix}. 
Its parametric value is given by
\begin{equation}
\label{eq:k1}
k_1 \sim \alpha' \Lambda_{TC}^{\gamma_\cU}  \left(\frac{\Lambda_{TC}}{\hoch}\right)^2  \left(  \frac{\Lambda_\cU}{\Lambda_{TC}} \right)^{\gamma_\cU} \Cw{\gamma_{\cal T}} \, , \qquad 
\Cw{\phantom{x}} = \left( \frac{\Lambda_{\cal T}}{\Lambda_{TC}}   \right)
\end{equation}
which we have normalized to the TC 
scale

The value of $k_1$ is, of course, suppressed by
the large scale $\hoch$ per se, 
but receives enhancements
from the powers of the anomalous dimensions.
For the maximal allowed anomalous dimensions 
 $\gamma_\cU \simeq \gamma_{\cal T} \simeq 2$ and a hierarchy 
of scales as indicated in the caption 
of Fig.~\ref{fig:three} one finds
$k_1 \Lambda_{TC}^{-\gamma_\cU}  \simeq \alpha' \cdot  O(10^{-2})$. We therefore  
expect the coupling $g_{h {\cal O}_\cU}\Lambda_{TC}^{-\gamma_\cU
} $ \eqref{eq:Lmix} to 
be considerably smaller than one. 

In this case there is generally a unique solution 
to the pole equation.
{In the IR region $1 < \dU  < 2$ the analysis can 
be made quantitative whereas in the UV region
 $2 < \dU  < 3$ the uncertainty due
to sizable counterterms
makes a quantitative assessment 
 difficult. 
 As explained in subsection  \ref{sec:UV}
 these counterterms are expected to be 
 comparable in size only for
 the leading coefficient $a_0$ \eqref{eq:as}.}

At the qualitative level 
it is an interesting question of whether the Higgs resonance is below or above the threshold \cite{van der Bij:2006pg,Delgado:2007dx}.
{For the values chosen 
in the caption of Fig.~\ref{fig:three} the
Higgs resonance is close to the IR cut off. 
On the other hand it could very well be that
the scale $\hoch$ is closer to the 
GUT scale which would decrease the IR cut off
definitely below the composite Higgs mass scale.} 
In appendix \ref{app:heidi} we comment
on the unparticle limit of the HEIDI models 
and compare our parameters with the fit
of that model 
to the excess of the Higgs search at LEP.

\section{Outlook}
\label{sec:outlook}

We introduced a framework  in which the Higgs and the unparticle are both composite. The underlying theories are four dimensional, asymptotically free,
nonsupersymmetric gauge theories with fermionic matter.  We sketched a
possible unification of these two sectors at a scale much higher than the electroweak scale. The resulting model resembles
extended technicolor models and we termed it extended technicolor unparticle (ETU). The coupling of the unparticle sector to the SM emerges in a simple way and assumes the form of  four-Fermi interactions  below
$\hoch$. 
 
In our model  the unparticle sector
is coupled  to the composite Higgs. Another possibility
is to assume that the Higgs sector itself is unparticle-like, with a continuous mass distribution.
This UnHiggs \cite{Stancato:2008mp,Calmet:2008xe}
could find a natural setting within
walking technicolor, which is part of our framework.
Of course it is also possible to think of an unparticle 
scenario that is not coupled to the electroweak sector,
where  scale invariance is broken at a (much) lower scale.
This could result in  interesting effects on low energy
physics as extensively studied in the literature.

With respect to our model in the future one can:

\begin{itemize}

\item{Study the composite Higgs production  in association with a SM gauge boson, 
both for proton-proton (LHC) and proton-antiproton 
(Tevatron) collisions via the low energy effective theory
\eqref{eq:sigmaeff}.  
In references 
\cite{Belyaev:2008yj,Zerwekh:2005wh} it has been demonstrated that such a process is enhanced with respect to the SM, due to the presence of a light composite (techni)axial resonance\footnote{
A similar analysis within an  extradimensional set up has been performed in \cite{Agashe:2008jb}.}. 
The mixing of the light composite Higgs with the unparticle sector modifies these processes in a way that can be explored at colliders.
Concretely, the transverse missing energy spectrum can be used to disentangle the unparticle sector from the TC contribution per se.}

\item{Use first principle lattice simulations 
to gain insight on the nonperturbative (near) conformal dynamics. 
It is clear from our analysis that this knowledge is crucial for describing and understanding unparticle dynamics.
As a model example we have considered in the main text partially gauge technicolor introduced  in \cite{Dietrich:2005jn}. These gauge theories are being studied on the lattice \cite{Appelquist:2007hu,Deuzeman:2008sc,Fodor:2008hn}.
Once the presence of a fixed point is established,
for example via lattice simulations \cite{Shamir:2008pb,Svetitsky:2008bw,DeGrand:2008dh,Fodor:2008hm}, 
 the anomalous dimension of the 
fermion mass can be determined from the
conjectured all order beta function 
\cite{Sannino:2008ha,Ryttov:2007cx}, as done
in section \ref{sec:ETU}.
Moreover, on the lattice one should be able to  
directly  investigate the two-point function, i.e. the unparticle propagator.}

\item{Investigate different models at the ETU level.  For example one could adapt some models,
introduced to generate masses to the SM fermions,
in  \cite{Christensen:2006rf,Christensen:2005bt,Appelquist:2004df,Appelquist:2004ai,Appelquist:2003hn,Appelquist:2003uu,Gudnason:2006mk} to  
improve on our ETU model.}

\item{Study possible cosmological consequences of our framework. The lightest baryon of the unparticle
gauge theory, the {\it Unbaryon}, is naturally stable (due to a  protected $U(1)$ unbaryon number) 
and therefore a possible dark matter candidate.
Due to the fact that we expect a closely spaced spectrum of Unbaryons and 
unparticle vector mesons, 
it shares properties in common with secluded models of dark matter \cite{Pospelov:2007mp} or previously discussed unparticle dark matter models \cite{Kikuchi:2007az}.}
\end{itemize}

Within our framework unparticle physics emerges as a natural extension of dynamical models
of electroweak symmetry breaking.  As seen above the link opens the doors to yet unexplored collider 
phenomenology and possible new avenues 
for dark matter, such as the use of the 
Unbaryon.

\section*{Acknowledgments}
FS thanks Mads T. Frandsen for discussions. RZ is grateful to Johan Bijnens, Oliver Brein, Mike Pennington, Tilman Plehn and Thomas Gregoire
for discussions.  
The work of FS is supported by the Marie Curie Excellence Grant under contract MEXT-CT-2004-013510.  RZ gratefully acknowledges the
support of an advanced STFC fellowship.
Further support is offered by the 
Marie Curie research training networks contract Nos\
MRTN-CT-2006-035482, {\sc Flavianet}, and MRTN-CT-2006-035505, {\sc Heptools}.

\appendix
\setcounter{equation}{0}
\renewcommand{\theequation}{A.\arabic{equation}}

\section{$\gamma$-induced condensate $\vev{{\cal O}_\cU}$}
\label{app:NJL}

In this appendix we  intend to sketch how
the $\gamma$-term, in addition 
to the $\alpha$-term in Eq.~\eqref{eq:unscen}, can induce 
induce an  unparticle VEV. 
The treatment essentially follows 
the Nambu$-$Jona-Lasino model 
\cite{Nambu:1961tp}; a simple and concise summary of the latter is given in the appendix of reference
\cite{Hill:2002ap}. The $\gamma$-term
\begin{equation}
{\delta \cal L }_{\Lambda_\cU}^{\rm eff} = \gamma' \frac{ \Lambda_\cU^{2 \gamma_\cU}}{\Hoch{2}}  {\cal O}_\cU ^2 
\end{equation}
can be rewritten into the following form
\begin{equation}
\delta {\cal L}^{\rm eff}_{\Lambda_\cU}  =  
(\sqrt{\gamma'} \frac{\Lambda_\cU}{\hoch} \Lambda_\cU^{\gamma_\cU} {\cal O}_\cU H + h.c)  - \Lambda_\cU ^2 |H|^2 \,,
\end{equation}
by the purely formal manipulation of introducing an auxiliary field $H$.
The crucial question 
is then whether the coupling of the $\gamma$-term is large enough to 
enforce a dynamical VEV.
This will be decided solely by the sign of the $|H|^2$-term. 
One has to integrate out the fermions between
the scales $\Lambda_\cU$ and $\mu$. 
This is straightforward in the unparticle scenario since the propagator is 
known, up to UV and IR cut offs.
The  $|H|^2$-term is then
simply given by contracting the 
unparticle propagator \eqref{eq:prop2}
between two  $ {\cal O}_\cU H$ interaction points; 
this leads to
\begin{equation}
\delta {\cal L}^{\rm eff}_{\mu} = - \Lambda_{\cU}^2 |H|^2 
\Big( 1 + \gamma' \frac{\Lambda_\cU^{2 \gamma_\cU  }}{\Hoch{2}}
\Delta_\cU(-\mu^2,\Lambda_{\rm UV}^2,\Lambda_{\rm IR}^2) \Big) \equiv -{m_H^2} |H|^2   \, .
\end{equation}
For $m_H^2 < 0$ the $|H|$ field acquires a VEV and induces an unparticle VEV through the gap equation. 
We remind the reader that 
the value of $\gamma'$ is expected to be of the order one.
In the range $\gamma_\cU \sim 0-1$ the sign of $m_H^2$
is negative for $\gamma' > \Hoch{2}/\Lambda_\cU^2$ which 
would demand an unnatural enhancement of the $\gamma'$
coefficient. For $\gamma_\cU > 1 $ the
$\gamma$-term becomes a relevant operator 
and one could expect a  qualitatively change in the picture. 
Around $\gamma_\cU \lesssim 2$ the inequality becomes
$\gamma' > (\Lambda_{\cal T}/\Lambda_\cU)^{2/3}
(\hoch/\Lambda_\cU)^{2/3} (\Lambda_{TC}/\Lambda_\cU)^{4/3}$ and could indeed lead to VEV at a scale comparable 
to the one from the $\alpha$ VEV \eqref{eq:unVEV}. 
To determine the value of the VEV we would need to evaluate
the coefficient of the $|H|^4$-term which is a difficult task
per se and beyond the scope of this paper.

\section{Comparison with Na\"ive Dimensional Analysis}
\label{app:NDA}

In reference \cite{Fox:2007sy}  it was pointed out that 
the interaction of a SM operator to the unparticle sector 
would act as a source of breaking the scale invariance. 
In the absence of an underlying model, 
the authors resorted to NDA. We will see here that the physics of the condensate or the anomalous dimension is, of course, not captured by 
such a generic approach as in reference \cite{Fox:2007sy}.

The schematic notation in \eqref{eq:unscen} made more
precise \cite{Georgi:2007ek} reads,
\begin{equation}
\label{eq:prec}
{\cal L}^{\rm eff} =  \Hoch{4} \, \Big(\frac{\Lambda_\cU}{\hoch}\Big)^{
d_{\rm UV} - \dU} \; \Big(
\frac{O_{\rm SM}}{\Hoch{d_{\rm SM}}}  \Big)
\Big(    \frac{{\cal O}_\cU}{\Hoch{\dU}} \Big) \,,
\end{equation}
where $d_{\rm SM/UV}$ are the scaling dimensions of the SM operator and the unparticle operator in the
UV.
Assuming that $O_{\rm SM} \to v^{d_{\rm SM}}$ acquires  a VEV at the electroweak 
scale $v$, NDA then suggests that scale invariance is broken at
a scale $\Lambda_{\rm IR}^{\rm NDA}$,
\begin{equation}
\label{eq:NDA}
(\Lambda_{\rm IR}^{\rm NDA})^ 4 \simeq {\cal L}^{\rm eff}({\cal O}_\cU \to (\Lambda_{\rm IR}^{\rm NDA})^\dU,O_{\rm SM} \to v^{d_{\rm SM}}) \,,
\end{equation}
when the term in Eq.~\eqref{eq:prec} is of the same size as  a generic four dimensional operator of the unparticle sector. This leads to 
\begin{equation}
\Lambda^{\rm NDA}_{\rm IR} \sim  v \Big(\frac{\Lambda_\cU}{\hoch}\Big)^{
\frac{d_{\rm UV} - \dU}{4 - \dU}} 
\Big(\frac{v}{\hoch}\Big)^{
\frac{d_{\rm SM}}{4 - \dU}-1}  \,.
\end{equation}
The above equation reduces to (3.3) in reference \cite{Fox:2007sy} 
for   $O_{\rm SM} = |H|^2$ with $d_{\rm SM} = 2$.

In our work 
$O_{\rm SM} \to O_{\rm TC} = \bar Q Q$ \eqref{eq:scem} with 
$d_{\rm SM} = 3$. The role of the electroweak scale is taken
by $v \to \Lambda_{\rm TC}$. 
The knowledge of the UV completion
settles the question on the UV dimension; 
$d_{\rm UV} = 3$.  Furthermore the
anomalous dimension $\gamma_{\cal T}$ introduces an additional multiplicative factor 
$\Cw{\gamma_{\cal T}}$ \eqref{eq:OUint} to the 
Lagrangian density \eqref{eq:prec} as an artefact
of walking technicolor.
Altogether this yields
\begin{equation}
 \Lambda^{\rm NDA}_{\rm IR}  \to
\Lambda_{\rm TC} \, (\cw)^{\frac{\gamma_{\cal T}}{1+\gamma_\cU}} 
\Big(\frac{\Lambda_\cU}{\hoch}\Big)^{\frac{\gamma_\cU}{1+\gamma_\cU}} 
\Big(\frac{\Lambda_{\rm TC}}{\hoch}\Big)^{\frac{2-\gamma_\cU}{1+ \gamma_\cU}} \, .
\end{equation}
The crucial question is then how this compares with 
the IR cut off in \eqref{eq:IR}. 
We find that, for generic values of  
the parameters, 
the condensate IR cut off is
\begin{equation}
\label{eq:IRcompA}
\Lambda_{\rm IR}  \gtrsim     \Lambda^{\rm NDA}_{\rm IR} \, ,
\end{equation}
larger than the IR cut off suggested 
by NDA.
 The essential point is that the VEV is sensitive to the UV cut off for $\dU \geq 2$ through the spectral integral \eqref{eq:Sreg}, 
whereas there is no such notion in the NDA. In fact in the NDA, c.f. Eq.~\eqref{eq:NDA}, 
it is built in that the unparticle operator assumes the IR-scale $\Lambda^{\rm NDA}_{\rm IR}$. 

This suggests that parametric equality \eqref{eq:IRcompA} is reached in the IR sensitive domain 
$\dU \gtrsim  1$. 
Most reassuringly it is verified that in this domain 
both IR cut offs scale as $\Lambda_{\rm TC} (\Lambda_\cU/\hoch)^{2/3}$. In the UV domain
$\dU \lesssim 3$ the scaling differs, 
$\Lambda_{\rm IR}^{\rm NDA} \sim \Lambda_{\rm TC} (\Lambda_{\rm TC}/\hoch)^2$ and 
$\Lambda_{\rm IR} \sim  \Lambda_{\rm TC} (\Lambda_\cU/\hoch)^{2/3}$.
In Fig.~\ref{fig:three}(right)  
the logarithm of the ratio of the two IR cut offs is plotted against $\gamma_\cU$ for specified input values and provides an example of the 
qualitative statement made above.

\section{Normalization factor $B_\dU$} 
\label{app:BdU} 
 
\begin{figure}
\begin{center}
    \includegraphics[width=0.4\textwidth]{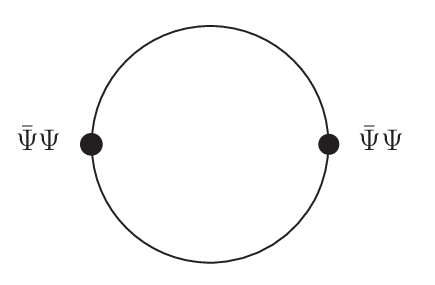}
  \end{center}
  \vspace{-20pt}
  \caption{\small Fermion 
bubble with scalar vertices, corresponding to the  unparticle propagator in the limt $\dU \to 3$.}
  \vspace{-5pt}
  \label{fig:bubble}
\end{figure}

In this appendix  we shall 
discuss the normalization factor 
$b_\dU \equiv \sqrt{B_\dU/(2\pi)}$  (\ref{eq:adu}) and  (\ref{eq:bdu}).
This is a necessary task in order
to extract quantitative results from the unparticle VEV equation \eqref{eq:unVEV}. 
Generally we do not know the behaviour of $B_\dU$ as a function of $\dU \equiv  3 - \gamma_\cU$, except around 
$\dU \gtrsim 1$ and at $\dU = 3$. 
Firstly, it is a fact that at $\dU = 1$
the operator ${\cal O}_\cU$ is equivalent to a free field
\cite{Mack:1975je}. This fixes the normalization factor, 
\begin{equation}
\label{eq:adu1}
B_\dU = 2 \pi (\dU -1) +  {\cal O } ( (\dU-1)^\eta  )  \quad {\rm with} \;\; \eta > 1 \,, {\rm for} \;
\; \dU \gtrsim 1  \,,
\end{equation}
around $\dU \gtrsim 1$ in a model independent way.
Since with \eqref{eq:adu1} 
the  spectral function \eqref{eq:adu} precisely produces the free massless field limit,
\begin{equation}
\lim_{\dU \to 1^+} \rho_\dU(s)  =  \lim_{\dU \to 1^+}
 (\dU -1)  s^{\dU-2} \theta(s)  = \delta(s) \,,
\end{equation}
with unit residue. This is equivalent to 
Georgi's \cite{Georgi:2007ek} requirement that $A_\dU$, in the notation of that paper, has to reproduce
the 1-particle phase space in that limit. 
Secondly, in our model at $\dU = 3$ the fermions
are free fields and the unparticle propagator, which 
we write in a K\"all\'en-Lehmann form,
\begin{equation}
\label{eq:prop}
\Delta_{\cU}(q^2) \equiv -i \int \! d^4 x e^{i \, q\cdot x}
\matel{0}{T \,  {\cal O}_\cU (x)  {\cal O}_\cU^\dagger (0)}{0}
= - \int \frac{ds \, \rho_\dU(s)}{s-q^2-i0} +  s.t. 
  \, ,
\end{equation}
has to reduce to the free fermion loop
depicted in Fig.~\ref{fig:bubble}. The letters $s.t.$ denote 
possible subtraction terms which are relevant for $\dU > 2$ to be discussed in subsection \ref{sec:regProp}.
This fixes the spectral function or  the normalization factor $B_\dU$ \eqref{eq:adu} at $\dU=3$ to
\begin{equation}
\label{eq:B3}
\rho_3(s) = s \frac{N_U }{8 \pi^2}   \qquad\leftrightarrow \qquad  B_3 = \frac{N_U }{4 \pi} \, .
\end{equation}
This value is different from 
\begin{equation}
   A_3 = 1/(256 \pi^3) 
\end{equation}
obtained from the normalization,
\begin{eqnarray}
\label{eq:aduGeorgi}
A_\dU  =  \frac{16 \pi^{5/2}}{(2\pi)^{2\dU}} 
\frac{\Gamma(\dU+1/2)}{\Gamma(\dU-1) 
\Gamma(2\dU)} 
= \frac{1}{2}\frac{1}{(4 \pi)^{2 \dU-3}}\frac{1}{\Gamma(\dU)\Gamma(\dU-1)}  
\end{eqnarray}
in reference \cite{Georgi:2007ek}. This is not surprising since
in this reference it was proposed to adapt
$A_\dU$ as the analytic continuation 
of the phase space of an integer number of $\dU$ massless 
particles. The operator 
${\cal O}_\cU = \varphi_0^\dU$, with $\varphi_0$ denoting a free massless scalar field, 
is of course a special realization of the
unparticle scenario for integer scaling dimension $\dU$.
We would like to emphasize that in reference 
 \cite{Georgi:2007ek} it was 
clearly stated that the actual normalization might be rather different from the 
one in a concrete model.

In the case at hand ${\cal O}_\cU|_{\dU = 3} = \bar \Psi_0
\Psi_0$ corresponds to two free fermions, instead of 
three free boson, which explains the difference. 
One could in principle generalize this scenario 
to higher powers of pairs of free fermion fields 
and adapt it as the normalization conditions for 
$B_\dU$ via analytic continuation. 
Unfortunately it appears that
no closed formula can be written down for this case. In order to obtain 
some quantitative results we resort to model $B_\dU$ by a quadratic interpolation function,
\begin{equation}
\label{eq:interpolation2}
(B_\dU)_{\rm interpol} = 2 \pi (\dU -1) + 
\Big( \frac{N_U}{16 \pi}-\pi \Big) (\dU - 1)^2 \, .
\end{equation}
Please not that these interpolation formula is 
positive as required by a positive spectral function 
\eqref{eq:adu}.
We would like to emphasize once more that the only firmly known parts are $B_3$ (model dependent) and the behaviour around $B_1$ (model independent).

\section{Unparticle limit of HEIDI models}
\label{app:heidi}

In the  HEIDI model 
\cite{vanderBij:2007um} the Higgs-Higgs propagator 
assumes the 
\begin{equation}
\label{eq:vdBij}
\Delta_{hh}(q^2) = \frac{1}{q^2-M^2 - c^2
(m_{0}^2-q^2-i0)^{\frac{d-6}{2}}} \, .
\end{equation}
same form as in \eqref{eq:higgsprop}.
The letter $c$ denotes a 
dimensionful constant proportional to the mixing parameter, $m_{0}$ is the mass of 
the lowest Kaluza Klein excitations and $M$ is the tree-level
Higgs mass. 
Comparing with \eqref{eq:higgsprop} 
and \eqref{eq:vdBij} it is readily seen that
identifying $(d-6)/2 = \dU -2$ and $m_0$ with 
$\Lambda_{IR}$ leads to a qualitatively similar
propagator.

The authors in reference \cite{vanderBij:2007um} attempted 
to reproduce the strictly scale invariant 
unparticle propagator in Eq.~\eqref{eq:georgi} 
\cite{Georgi:2007si}. This can be achieved 
by making the mixing arbitrarily 
large  $c  \gg 1$, adopting $m_0 \to 0$ and keeping
$M$ fixed. The Higgs-Higgs propagator then 
becomes the inverse unparticle 
propagator, which in turn 
looks like an unparticle propagator with reversed scaling power. On this basis the identification 
$(d-6)/2 = -(\dU -2) \rightarrow  \dU = 5-d/2$ 
was proposed in 
\cite{vanderBij:2007um}.
In our model the unparticle itself couples to
the techniquark or technihadrons directly and
such a limit  does therefore not seem
necessary for unparticle-like effects in phenomenology.

Finally, the excess of the LEP data 
in the Higgs strahlung search
of $2.3\sigma$ and $1.7\sigma$ 
around $98\,{\rm GeV}$ and $115\,{\rm GeV}$ 
respectively were fitted to the HEIDI model \cite{vanderBij:2007um}. 
The first peak is interpreted as the Higgs resonance and the second one as the onset of the continuum.
Comparing the fitted parameters in \cite{van der Bij:2006pg} 
with $k_1$ in Eq.~\eqref{eq:k1} 
and input values as given in the caption of 
Fig.~\ref{fig:three} it is seen that they are
of the same order of magnitude.

\end{document}